\documentclass[aps,pra,showpacs,twocolumn]{revtex4}
\usepackage{graphicx}
\usepackage{amsmath,amssymb,amsthm,dsfont,bm}

\begin{document}
\preprint{}
\title{All states are nonclassical: entanglement of joint statistics}
\author{Alfredo Luis} 
\email{alluis@fis.ucm.es}
\homepage{http://www.ucm.es/info/gioq}
\affiliation{Departamento de \'{O}ptica, Facultad de Ciencias
F\'{\i}sicas, Universidad Complutense, 28040 Madrid, Spain}
\date{\today}

\begin{abstract}
Joint measurements of two observables reveal that every state is nonclassical, with the only trivial exception of 
the state with density matrix proportional to the identity.  This naturally includes states considered universally as 
classical-like, such as SU(2) and Glauber coherent states.  We show that this holds because we can always find 
a joint measurement whose statistics is not separable. 
\end{abstract}

\pacs{03.65.Ta, 42.50.Dv, 42.50.Xa }

\maketitle

\noindent  {\em Introduction.--} 
The proper understanding of nonclassical effects is naturally a crucial issue of the quantum theory, actually its 
{\em raison d'\^ etre}. But even beyond that, quantumness is the resource behind the quantum-technology revolution. 

In any sounded approach to the subject,  the conclusion is that nonclassical effects always emerge as the 
impossibility of confining randomness of two or more variables within probability distributions  \cite{thisone}. 
For example, this is actually the case of the celebrated quantum tests of the Bell type \cite{WMB,AF,TMN}.  
This includes as a particular case mainstream tests  such as the failure of the Glauber-Sudarshan $P$ 
function to be a true probability density \cite{P}. 

In this work we show from scratch that every state is nonclassical, with the only trivial exception  
of the state with density matrix proportional to the identity. This naturally includes states universally considered 
as classical-like, such as SU(2) and Glauber coherent states and, accordingly, all cases where the $P$ function 
is a true probability distribution. 

We show that these results hold via entanglement of joint statistics of two observables measured simultaneously.
To this end the measurement takes place in an enlarged space. Nonclassicality manifests when showing that 
there are joint measurements of pairs of observables whose random outputs  cannot be described by a {\it  bona 
fide} joint distribution of system random variables. To this end, entanglement of joint statistics is crucial since 
the observed statistics is always separable in classical physics. 

There are several arguments that highlight the relevance of these results and the way they may open a new area 
of research:  This provides a universal route to nonclassical effects, embracing all known to date as well as any 
other to be discovered in the future. In particular it shows that we can find nonclassical effects anywhere. 
Moreover, it suggests that we can achieve nonclassical effects via detection as well as via state preparation. 
This formulation draws many parallels with Bell-type measurements.  As a  matter of fact,  entanglement reveals 
to be a crucial property, but this time the entanglement refer to statistics instead of particles. Being defined in 
measurements actually performed the approach presented here is practical from the very beginning. 

\bigskip

\noindent  {\em Basic settings.--}
Nonclassical effects can only emerge when addressing the joint statistics of multiple observables,  
specially if they are incompatible. Nonclassicality cannot be a single-observable property since 
within classical physics it is always possible to reproduce exactly the statistics of any quantum 
observable. Note that this multi-observable nature is the common denominator of the most renowned 
landmarks of the quantum theory, such as complementarity, uncertainty relations and Bell tests \cite{WMM}. 

In the most general case, joint measurements require the coupling of the system space ${\cal H}_s$ 
with auxiliary degrees of freedom ${\cal H}_a$. We consider the simultaneous measurement of two 
compatible observables, $\tilde{X}$ and $\tilde{Y}$, in the enlarged space ${\cal H}_s \otimes {\cal H}_a$  
with outcomes $x$ and $y$, respectively, and joint probability $\tilde{p}_{X,Y} (x,y)$. Since this corresponds 
to the statistics or a real measurement we have that  $\tilde{p}_{X,Y} (x,y)$ is a well-behaved probability 
distribution. The corresponding marginal distributions are 
\begin{equation}
\label{itn}
\tilde{p}_X (x)  = \sum_y \tilde{p}_{X,Y} (x,y), \quad \tilde{p}_Y (y) = \sum_x \tilde{p}_{X,Y} (x,y) ,
\end{equation}
where we are assuming a discrete range for  $x$ and $y$ without loss  of generality. We can always 
consider that these marginals provide complete information about two system observables in the system 
space ${\cal H}_s$, say $X$ and $Y$, respectively. This is to say that their probability distributions 
$p_A  (a)$ for $A=X,Y$ and $a, a^\prime=x,y$ can be retrieved from the observed marginals  $\tilde{p}_A  (a)$ as
\begin{equation}
\label{inv}
p_A (a) = \sum_{a^\prime} \mu_A (a, a^\prime ) \tilde{p}_A  (a^\prime) ,
\end{equation}
where the functions $\mu_A (a, a^\prime )$ is completely known as far as we know the measurement being 
performed and the initial state of the auxiliary degrees of freedom ${\cal H}_a$. The key idea is to extend 
this inversion (\ref{inv}) from the marginals to the complete joint distribution to obtain a joint distribution 
$p_{X,Y} (x,y)$ for the $X,Y$ variables in the system state \cite{WMM}
\begin{equation}
\label{nti}
p_{X,Y} (x,y)  = \sum_{x^\prime, y^\prime} \mu_X (x, x^\prime) \mu_{Y} (y, y^\prime )  
\tilde{p}_{X,Y} (x^\prime , y^\prime ) .
\end{equation}
Parallels can be drawn with the construction joint probability distributions via the inversion of moments \cite{UD}. 

\bigskip

\noindent  {\em Classical physics.--}
Let us show that in classical physics this inversion procedure (\ref{nti})  always leads to a \textit{bona fide} probability 
distribution $p_{X,Y} (x,y)$. Classically the state of the system can be completely described by a legitimate probability 
distribution  $p_j$, where index $j$ runs over all admissible states $\bm{\lambda}_j$ for the system. This is the corresponding 
phase space, that we will represent by three-dimensional real vectors $\bm{\lambda}_j$, assumed to form a discrete 
set for simplicity and without loss of generality.  There is no limit to the number of vectors $\bm{\lambda}_j$ so it may approach 
a continuum if necessary. 

So the observed joint statistics can be always expressed as
\begin{equation}
\label{jj}
\tilde{p}_{X,Y} (x,y) = \sum_j p_j \tilde{X} (x | \bm{\lambda}_j)  \tilde{Y} (y | \bm{\lambda}_j) ,
\end{equation}
where  $\tilde{A}(a | \bm{\lambda}_j) $ is the conditional probability that the observable $\tilde{A}$ 
takes the value $a$ when the system state is $\bm{\lambda}_j$. By definition, phase-space points $\bm{\lambda}_j$ 
have definite values for every observable so the factorization $\tilde{X} (x | \bm{\lambda}_j)  \tilde{Y} 
(y | \bm{\lambda}_j)$ holds, strictly speaking as the product of delta functions. However, it is simpler to refer to 
conditional probabilities in general. Applying Eq. (\ref{inv}) we get the conditional probabilities for the system variables  
\begin{equation}
\label{rel}
 A( a | \lambda_j ) = \sum_{a^\prime} \mu_A (a,a^\prime )  \tilde{A} ( a^\prime | \bm{\lambda}_j ) .
\end{equation}
Thus, because of the separable form (\ref{jj}) we readily get from Eqs. (\ref{nti}) and (\ref{rel}) that the result of the inversion 
is  the actual joint distribution for $X$ and $Y$
\begin{equation}
p_{X,Y} (x,y)  = \sum_j  p_j X(x | \bm{\lambda}_j)  Y (y | \bm{\lambda}_j ) ,
\end{equation}
and therefore a legitimate statistics. Thus, lack of positivity or any other pathology of the retrieved joint distribution $p_{X,Y} (x,y) $ 
is then a signature of nonclassical behavior. 

\bigskip

\noindent {\em Qubit example.-- }
Let us focus on the qubit as the simplest quantum system ${\cal H}_s$, with the aim that the results readily extend to 
any system space. The most general state of the qubit is 
\begin{equation}
\label{ss}
\rho = \frac{1}{2} \left ( \sigma_0 + \bm{s} \cdot \bm{\sigma}  \right ) , \quad |\bm{s} | \leq 1, 
\end{equation}
where $\bm{s}$ is a three-dimensional real vector with $| \bm{s}| \leq 1$, $\sigma_0$ is the identity, and $\bm{\sigma}$ 
are the Pauli matrices. The task is finding for every $\rho$ a suitable measurement so that the inversion $p_{X,Y} (x,y)$  
of the observed statistics  $\tilde{p}_{X,Y} (x,y)$ cannot be a probability distribution. To this end, we will use that any 
measurement performed in the enlarged space ${\cal H}_s  \otimes {\cal H}_a$ can be conveniently described by a positive 
operator-valued measure in ${\cal H}_s$ 
\begin{equation}
\label{tpovm}
\tilde{\Delta}_{X,Y} (x,y) = \frac{1}{4} \left ( \sigma_0 + \bm{\eta} (x,y) \cdot \bm{\sigma}  \right )  .
\end{equation}
Positivity and normalization require that 
\begin{equation}
\tilde{\Delta}_{X,Y} (x,y) \geq 0, \qquad \sum_{x,y} \tilde{\Delta}_{X,Y} (x,y) = \sigma_0 ,
\end{equation}
so that 
\begin{equation}
| \bm{\eta} (x,y) |  \leq 1 , \qquad \sum_{x,y}  \bm{\eta} (x,y) = \bm{0} .
\end{equation}
The corresponding statistics is 
\begin{equation}
\tilde{p}_{X,Y} (x,y)= \mathrm{tr} \left [ \rho \tilde{\Delta}_{X,Y} (x,y) \right ] = \frac{1}{4} \left ( 1 +  \bm{\eta} (x,y) \cdot \bm{s}  \right ) ,
\end{equation}
and naturally 
\begin{equation}
\tilde{p}_{X,Y} (x,y) \geq 0, \qquad  \sum_{x,y} \tilde{p}_{X,Y} (x,y) = 1.
\end{equation}

For definiteness, let us consider the case  
\begin{equation}
\bm{\eta} (x,y) = \frac{\eta}{\sqrt{3}} \left (x, y,xy \right ) ,
\end{equation}
where  $x,y = \pm 1$ and  $\eta$ is a real parameter we will assume positive without loss of generality $1 \geq \eta >0$.
Actually, for $\eta = 1$ we have that $\tilde{p}_{X,Y} (x,y)$ is a discrete and complete sampling of the SU(2) Husimi function 
for two-dimensional systems \cite{dQ}. The observed marginals are
\begin{equation}
\label{oe}
\tilde{p}_X (x)  = \frac{1}{2} \left ( 1 +  x \frac{\eta}{\sqrt{3}} s_x \right ),
\quad
\tilde{p}_Y (y)  = \frac{1}{2} \left ( 1 +  y \frac{\eta}{\sqrt{3}} s_y \right ),
\end{equation}
that provide complete information about the system observables $X = \sigma_x$, $Y= \sigma_y$ with exact statistics 
\begin{equation}
p_X (x)  = \frac{1}{2} \left ( 1 +  x s_x \right ),
\qquad
p_Y (y)  = \frac{1}{2} \left ( 1 +  y s_y  \right ) .
\end{equation}
The inversion of the marginals is carried out by the functions
\begin{equation}
\mu_A \left ( a ,  a^\prime \right ) = \frac{1}{2} \left ( 1 + \frac{\sqrt{3}}{\eta}  a a^\prime \right )  ,
\end{equation}
so that the inversion of the joint distribution in Eq.  (\ref{nti}) leads to
\begin{equation}
p_{X,Y} (x,y) = \frac{1}{4} \left ( 1 + x s_x  +   y s_y + x y s_z \frac{\sqrt{3}}{\eta} \right ) .
\end{equation}

\bigskip

\noindent {\em Proof of principle.-- } 
Throughout we are free to chose the axes and the observables measured. In this spirit using SU(2) symmetry 
and without loss of generality we can choose axes so that $s_x = s_y =0, s_z = |\bm{s} |$, so that 
\begin{equation}
\label{es}
p_{X,Y} (x,y) = \frac{1}{4} \left ( 1 +   x y \frac{\sqrt{3}} {\eta}| \bm{s} |  \right ) .
\end{equation}
This can take negative values for $x = - y = \pm 1$ 
\begin{equation}
\label{neg}
p_{X,Y} (\pm 1,\mp 1) = \frac{1}{4} \left ( 1 -  \frac{\sqrt{3}} {\eta} | \bm{s} |  \right )  < 0,
\end{equation}
provided that $\eta < \sqrt{3}  | \bm{s} |$. Clearly for all $\bm{s} \neq \bm{0}$ we can always chose $\eta$ 
satisfying this relation. So every state different from the identity is non classical. 

In this regard it is worth noting that all pure states of the qubit are SU(2) coherent states \cite{cs}. 
These states, as well as all their incoherent superpositions, are reported  as classical according to 
their well-behaved Glauber-Sudarshan $P$-function. However we have just shown that  they are
actually nonclassical.

\bigskip

\noindent {\em Entanglement of statistics.-- } 
Let us provide an explicit demonstration that if $p_{X,Y}  <0$ the observed statistics (\ref{es}) cannot 
be expressed in a separable form 
\begin{equation}
\label{sep}
\tilde{p}_{X,Y} (x,y) = \sum_j  \frac{p_j}{4} \left ( 1 +  x \frac{\eta}{\sqrt{3}} \lambda_{j,x} \right )
\left ( 1 +  y \frac{\eta}{\sqrt{3}} \lambda_{j,y} \right ),
\end{equation}
leading to 
\begin{equation}
p_{X,Y} (x,y) = \sum_j  \frac{p_j}{4} \left ( 1 +  x \lambda_{j,x} \right )
\left ( 1 +  y  \lambda_{j,y} \right ),
\end{equation}
where $\lambda_{j,x}$ and $\lambda_{j,y}$ are the corresponding components of the vectors $\bm{\lambda}_j$,  
with $|\bm{\lambda}_j| \leq 1$. We recall that there is no limit to the number of vectors $\bm{\lambda}_j$. Note that 
in the separable  form (\ref{sep}) we have used the general form for the observed conditional probabilities 
$\tilde{A} ( a | \bm{\lambda}_j )$ as in Eq. (\ref{oe}). Then, if the factoriced form (\ref{sep}) holds we have for our 
case in Eq. (\ref{es}) 
\begin{equation}
\label{lc}
\sum_j p_j \lambda_{j,x} \lambda_{j,y} =   \frac{\sqrt{3}} {\eta}| \bm{s} |  . 
\end{equation}
We can readily show that separability (\ref{sep}) and negativity  (\ref{neg}) are contradictory. This  is because 
$|\bm{\lambda}_j| \leq 1$ so that $\sum_j p_j \lambda_{j,x} \lambda_{j,y} \leq 1$. Thus separability implies 
$\sqrt{3}  | \bm{s} |/\eta \leq 1$ while negativity just the opposite $\sqrt{3}  | \bm{s} |/\eta > 1$. Therefore, negativity 
of the inferred distribution $p_{X,Y} (x,y) $ is equivalent to entanglement of the observed statistics $\tilde{p}_{X,Y} (x,y)$. 

\bigskip

\noindent  {\em Conclussions and discussion.--}
We have shown that for every state, excluding the identity, we can devise a joint measurement of 
two observables so that the outcomes are incompatible with a {\em bona fide} joint probability distribution 
for system random variables. This impossibility is expressed by an inversion procedure that provides 
legitimate results for the marginal distributions, but leads to negative values for the joint distribution. 

We have illustrated this approach with a qubit system. The result can be extended to systems in Hilbert spaces of 
arbitrary dimension. For pure states $| \psi \rangle$ this can be readily done by focusing on the two-dimensional 
subspace spanned by the pair $| \psi \rangle$, $| \psi_\perp \rangle$, where $| \psi_\perp \rangle$ is any state 
orthogonal to $| \psi \rangle$. We may the define $\sigma_z = |\psi \rangle \langle  \psi |- | \psi_\perp \rangle \langle  
\psi_\perp |$ and accordingly for the other Pauli matrices. For mixed states we may we focus on their projection on 
any two-dimensional space that can be regarded as the marginal distribution of a larger statistics. Alternatively, we 
may deal with dichotomic observables, such as parity or any other on/off detectors \cite{par,SVA}. 

As a relevant consequence we have that this protocol discloses nonclassical effects for states customarily 
regarded as classical-like, or even as the most classical states. These are the Glauber coherent states and the  
SU(2) coherent states  for spin variables. Previous works have also reported nonclassical properties for these 
states following different approaches \cite{thisone,ncpn}, and fits with other results extending nonclassical
correlations and entanglement to all quantum states \cite{qq}. The case of he qubit is particularly interesting since 
all pure states are SU(2) coherent states so every state would be classical-like according to common understanding.
This is particularly meaningful in quantum optics where a qubit can be implement through the polarization of 
a single photon \cite{qp}, and therefore simply accessible to experiment.

We have shown that nonclassicality holds because the observed joint probability distribution becomes 
entangled. We have to stress that this entanglement does not refer to particles. Notice that all the analysis 
is made entirely within the system space ${\cal H}_s$ both for the system state (\ref{ss}) and the positive 
operator measure (\ref{tpovm}).  

In this regard, the complete picture bears a large relation to Bell-type measurements, specially by the failure of 
expression (\ref{jj}) \cite{WMB,AF}. So that the lack of factorization $\tilde{X} (x | \bm{\lambda}_j)  \tilde{Y} (y | 
\bm{\lambda}_j) $ can be interpreted as meaning the impossibility of assigning prescribed values to both variables 
independently. There are also similarities because in both scenarios incompatible observables are measured. 
There are also significant differences. For example, here all measurements are performed within a single arrangement 
instead of requiring multiple different settings. This is to say that in this case nonclassical effects cannot be jeopardized 
by lack of a common probability space between different settings \cite{WMB,TMN}. 
 
 \bigskip
 
\noindent  {\em Acknowledgements.--}
I thank Profs. A. R. Usha Devi, H. S. Karthik, A. Iosif, and M. Acedo for valuable comments, as well as encouraging 
support from Ceres team. I acknowledge financial support from Spanish Ministerio de Econom\'ia y Competitividad 
Project No. FIS2012-35583 and from the Comunidad Aut\'onoma de Madrid research  consortium QUITEMAD+ 
Grant No. S2013/ICE-2801.

\end{document}